\documentstyle[12pt,lnfprep,epsfig]{article}
\newcommand{\Header}{
  \begin{tabular}{rl}
  \hspace{-.4cm}\includegraphics{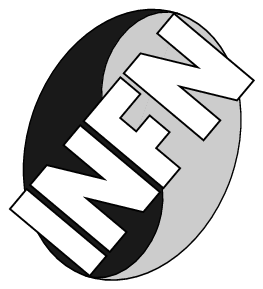} &
    \renewcommand{\arraystretch}{0.5}
    \begin{tabular}{r}
      {\hspace{1cm}~\LARGE\sffamily LABORATORI~ NAZIONALI~ DI~ FRASCATI}\\
      \\
      {\Large\sffamily SIS-Pubblicazioni}\\
    \end{tabular}
    \renewcommand{\arraystretch}{1}
  \end{tabular}
  \vskip 1cm
  \begin{flushright}
  \renewcommand{\arraystretch}{0.5}
    \begin{tabular}{r}
      {\underline{LNF-99/006(P)}}\\    
      {\small 12 Febbraio 1999} \\      
      \\
      {\small IISc/CTS/4/99} \\
      {\small\tt hep-ph/9903331} 
    \end{tabular}
  \end{flushright}
  \renewcommand{\arraystretch}{1}
  \vskip 1 cm
  }

\def\ra{\rightarrow}
\def\ben{\begin{subequations}}
\def\be{\begin{equation}}
\def\een{\end{subequations}}
\def\ee{\end{equation}}
\def\beq{\begin{eqalignno}}
\def\eeq{\end{eqalignno}}
\def\bea{\begin{eqnarray}}
\def\eea{\end{eqnarray}}
\def\qsq{\mbox{$Q^2$}}

\def\f2gam{\mbox{$ F_2^\gamma $}}

\def\rts{\mbox{$\sqrt {s}$}}
\def\pghad{\mbox{$P_{\gamma}^{\rm had}$}} 
\def\phad{\mbox{$P^{\rm had}$}}
\def\gamp{\mbox{$\gamma {\rm } p$}}
\def\gamgam{\mbox{$\gamma \gamma$}}
\def\eplem{\mbox{$e^+e^-$}}
\def\siggmhad{\mbox{$\sigma (\gamma \gamma \rightarrow {\rm hadrons})$}}
\begin{document}
\begin{titlepage}
\title{ 
  \Header
  {\large \bf Eikonalised minijet model predictions  for
cross-sections of photon induced processes }
}
\author{ Rohini M. Godbole$^1$, Giulia Pancheri$^2$\\
{\it ${}^{1)}$  CTS, Indian Institute of Science, Bangalore, 5600 012,
  India}
 \\
{\it ${}^{2)}$INFN, Laboratori Nazionali di Frascati, P.O. Box 13,
I-00044 Frascati, Italy}
} 
\maketitle
\baselineskip=14pt

\begin{abstract}
In this talk I  present the results of an analysis of total and 
inelastic hadronic cross-sections for photon induced processes
in the framework  of an eikonalised minijet model(EMM).  We fix the
various input parameters to the EMM calculations  by using the data
on photoproduction cross-sections and then make predictions for
\siggmhad\ using {\it same} values of the  parameters. We then compare
our predictions with the recent measurements of \siggmhad\ from LEP.
We also show  that in the framework of the EMM
the rise with \rts\ of the total/inelastic
cross-sections will be faster for photon induced processes than for 
the processes induced by hadrons like protons.
\end{abstract}

\vspace*{\stretch{2}}
\begin{flushleft}
  \vskip 2cm
{ PACS13.60.Hb,13.85.Lg} 
\end{flushleft}
\begin{center}

Presented  by the first author at the Workshop on photon 
interactions and the photon structure, in Lund, September
10-13, 1998.
\end{center}
\end{titlepage}
\pagestyle{plain}
\setcounter{page}2
\baselineskip=17pt

\section{Introduction}

The rise of hadronic total cross-sections  
$\sigma(A + B \ra {\rm hadrons})$ with energy, 
has been now observed for a set of comparable values of \rts\
where both A,B are hadrons~\cite{CDF,E710},
when one of them is a photon~\cite{HERAZ,HERAH1} and
when both of them are photons~\cite{L3,OPAL}.  It is well 
known that interactions of a photon with another hadron
or photon receive contributions from the `structure' of a photon
which the photon develops due to its fluctuation  into a virtual 
$q \bar q$ pair. The recent measurements~\cite{L3,OPAL} of 
\siggmhad\ at higher
energies of upto and beyond $\sim {\cal O} (100)$ GeV, have confirmed that the hadronic
cross-sections rise with \rts\ and preliminary claims~\cite{stefan,valeri} are that they rise faster than the $p \bar p$ cross-sections.
A measurement of $\gamma^* p$ cross-sections by ZEUS collaboration, extrapolated
to $\qsq\ = 0$ \cite{Bernds,talk},  lies above the 
photoproduction measurements~\cite{HERAZ,HERAH1}. These extrapolated
$\gamma^* p$ data and the new \gamgam\ data 
seem to indicate that the  rise of cross-sections with \rts\
gets faster as one replaces hadrons with photons successively.
In the Pomeron-Regge picture~\cite{DL} the total cross-section is 
given by 
\be
\sigma^{\rm tot}_{ab}=Y_{ab} s^{-\eta}+X_{ab}s^{\epsilon}
\label{regpom}
\ee
where $\eta$ and $\epsilon $  are related to the intercept at zero of the
leading Regge trajectory and of the Pomeron, respectively.  The value of
the Pomeron intercept indicated by the unpublished results of 
ZEUS~\cite{Bernds,talk} is $0.157\pm 0.019\pm0.04$ 
whereas the corresponding value for the \gamgam\ data obtained by the  L3 
collaboration~\cite{L3} 
is $0.158 \pm 0.006 \pm 0.028$ which is to be compared with the value
of $\sim 0.08$~\cite{DL} for pure hadronic cross-sections. 

\begin{figure}[htb]
\begin{center}
\mbox{\epsfig{file=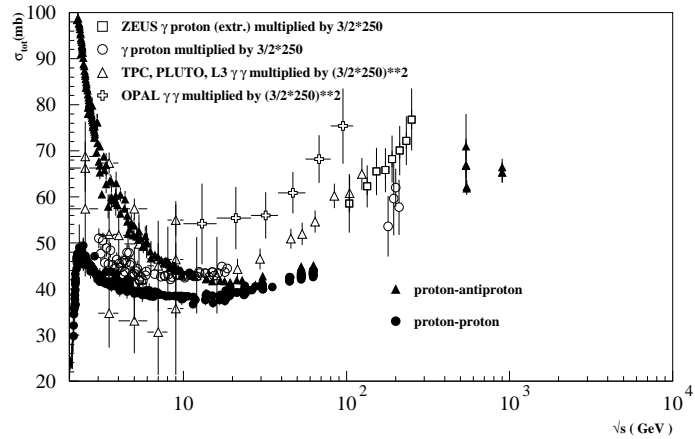,height=60mm}}
\end{center}
\caption{Energy dependence  of  $\sigma_{ab}^{\rm tot}$.
\label{figure1}}
\end{figure}

Fig.~\ref{figure1} shows the energy dependence of the
hadronic cross-sections as well as those for the photon
induced processes. The latter are multiplied by a quark model
motivated factor of $3/2$ and the inverse of the probability
for a photon to fluctuate into a $q \bar q$ pair: \pghad. The value 
chosen for \pghad\ is $1/250$ which is close to a value motivated by
VMD picture. i.e. 
\begin{equation}
\label{PVMD}
P_\gamma^{\rm had} = P_{VMD}=\sum_{V=\rho,\omega,\phi} {{4\pi 
\alpha_{\rm em}}
\over{f^2_V}}\approx {{1}\over{250}}.
\end{equation}
Fig.~\ref{figure1} illustrates  that {\it all} total cross-sections
show a similar rise in energy when the difference between photon and
hadron is taken into account, albeit with indications of somewhat
steeper energy dependence for photon induced processes\footnote{It 
should be noted here that although the rate of rise 
with \rts\ is similar for both OPAL and L3~\protect\cite{stefan}, 
on this plot the OPAL data seems to stand a bit apart, which may be
due to the difference in the normalisation of the two data sets.}.
The similarity in the energy dependence makes it interesting to 
attempt to give a description of all  three data sets in the same 
theoretical framework\cite{block}.  We have analysed the cross-sections 
for photon-induced processes alone ~\cite{photon97,fras,rohinipl}
and we find that the EMM calculations  generically predict faster 
rise with \rts\ for \gamgam\ case than would be expected by an universal 
pomeron hypothesis. I will further present arguments why in the framework 
of EMM the  Pomeron intercept is expected to increase as the number 
of colliding  photons in the process increase.

\section{Eikonalised Minijet Model}

There are some basic differences between the purely hadronic 
cross-section measurements and those of the total cross-sections in
photon induced processes. In the purely hadronic case the measurement
of the total cross-section comes from the  combined methods of
extrapolation of the
elastic diffraction peak and total event count,
 whereas in the case of  photon induced
processes, this has to be extracted from the data using a Montecarlo;
for example the total \gamgam\ cross-sections are  extracted 
from a measurement of hadron production in the untagged \eplem\ 
interactions. Experimentally this gives rise to different types of
uncertainites in the measurement\footnote{This is  clear from  
the discussions, for example, in Ref.~\protect\cite{stefan}.}. Theoretically, 
the concept of the `elastic' cross-section
can not be well defined for the photons. In models satisfying 
unitarity like those which use the eikonal formulation, it is 
important to understand what is the definition of the elastic 
cross-section and  if the data include all of this elastic
cross-section or part or a small fraction. To that end let us
summarise some of the basic features of the Eikonal formulation.

Let us start from the very beginning. Consider the eikonal 
formulation for the elastic scattering amplitude
\begin{equation}
f(\theta)={{i k}\over{2\pi}} \int d^2{\vec b} e^{i{\vec q}\cdot{\vec b}}
[1-e^{i\chi(b,s)/2}]
\label{elsc}
\end{equation}
which, together with the optical theorem leads to the expression for
the total cross-sections
\begin{eqnarray}
\sigma^{\rm el}&=&\int d^2{\vec b}
|1-e^{-i\chi(b,s)/2}|^2 \\
\label{eik1}
\sigma^{\rm tot}&=&2\int d^2{\vec b}
[1-e^{-\chi_I(b,s)/2}cos(\chi_R)] \\ 
\label{eik2}
\sigma^{\rm inel}&=&\sigma^{\rm tot}-\sigma^{\rm el}=\int d^2{\vec b}
[1-e^{-\chi_I(b,s)}]
\label{eik3}
\end{eqnarray}
According to the minijet model\cite{CLINE,minijet,eikminijets,treleani} 
the rise of total cross-sections can be calculated 
from QCD.
In the model, there is an {\it ad hoc} sharp division between 
a soft component, which is of non-perturbative origin and for which 
the  model is not able to make theoretical predictions, and a hard 
component, which receives input from perturbative QCD. 
The minijet model assumes that the  rise with energy of total cross-sections
is driven by the rise with energy of the number of low-x partons
(gluons) responsible for hadron collisions and in its  simplest formulation 
reads 
\begin{equation}
\sigma^{inel,u}_{ab}=\sigma_0 + \int_{p_{tmin}}
 d^2{\vec p_t} {{d\sigma_{ab}^{jet}}\over{d^2{\vec p_t}}}=\sigma_0+
\sigma^{jet}_{ab}(s,p_{tmin}),
\label{miniold}
\end{equation}
the superscript $u$ indicating that this is the ununitarised cross-section.
This concept can be embodied in a unitary formulation as in 
\ref{eik1}-\ref{eik3}, by writing \cite{FLETCHER,SARC}
\begin{equation}
\sigma^{inel}_{ab}= P^{\rm had}_{\rm ab}\int d^2{\vec b} [1-e^{-n(b,s)}]
\label{siginel}
\end{equation}
with
\begin{equation}
n(b,s)=A_{ab}(b)[\sigma_{h/a,h/b}^{soft}+
{{\sigma^{jet}_{ab}(s,p_{tmin})}\over{P^{\rm had}_{\rm ab}}}]
\label{numb}
\end{equation}
In eq.\ref{siginel}, we have inserted, to include the generalization 
to photon processes, a factor $P^{\rm had}_{\rm ab}$ defined as the probability that
particles $a$ and $b$ behave like hadrons in the collision. This 
 parameter is unity for hadron-hadron
processes, but of order $\alpha_{\rm em}$ or $\alpha_{\rm em}^2$ 
for processes with 
respectively one or two photons in the initial state. The definition of
$\sigma_{h/a,h/b}^{soft}$ in  eq.\ref{numb} is such that, even in the
photonic case, it is of hadronic size, just like 
${{\sigma^{\rm jet}_{\rm ab}(s,p_{tmin})}/{P^{\rm had}_{\rm ab}}}$.
A simple way to understand the need for this factor~\cite{ladinsky} is 
to realise that the unitarisation in this formalism is achieved by 
multiple parton  interactions in a given scatter of hadrons and once 
the photon has `hadronised' itself, one should not be paying the price 
of \pghad\ for further multiparton scatters.

At high energies, the dominant term in the eikonal is the $jet$
cross-section which is calculable in QCD and depends on the parton
densities in the colliding particles and $p_{tmin}$, which admittedly is
an ad hoc parameter separating the perturbative and nonperturbative 
contributions to the eikonal. The basic assumption in arriving at
eq.~\ref{siginel} is that the multiple parton scatters  responsible for
the unitarisation are independent of each other at a given value of $b$. 
In this model $n(b,s) $ in eq. \ref{numb}, is identified as the average 
number of collisions at any given energy \rts\ and impact parameter $b$. 
The $b$ dependence is assumed to be given by the function $A_{ab}(b)$ which
is modelled in different ways. This function measures the overlap of 
the partons in the two hadrons $a,b$ in the transverse plane. 

Before going to the discussion of different models of $A_{ab} (b)$, 
we note that the mini-jet model is particularly well suited for generalisation
to the photon-induced processes where the concept of `elastic' cross-section
is not very well defined. Whereas for the hadronic case one starts from the 
elastic amplitude followed by the optical theorem as done in 
eqs.~\ref{elsc} -- ~\ref{eik3}, in this case the starting point is
actually the eq.~\ref{siginel} and then one defines $\sigma^{\rm tot}_{ab}$
using  eq.~\ref{eik2} with $\chi_R = 0$ and using $\chi_I $ as given by 
\ref{siginel}. The above discussion specifies the total cross-section 
formulation of the EMM for photon-induced processes. While our
earlier analyses~\cite{photon97,fras,rohinipl} assumed that the 
\gamgam\ cross-sections presented were the inelastic cross-sections,
the analyis of \cite{block} had used the total cross-section formulation but
with a different ansatz for the eikonal. Our  analysis
uses the total cross-section formulation of the EMM with the perturbative part
of the Eikonal as given by QCD {\it a-la} eq.~\ref{siginel}. 

\section{Overlap function and jet cross-sections.}
The overlap function $A_{ab}(b)$ is normally calculated in terms of the 
convolution of  the matter distributions $\rho_{a,b} (\vec b)$of the 
partons in the colliding hadrons in the transverse plane 
\begin{equation}  
A_{ab}(b) =  \int d^2\vec{b'} \rho_a (\vec {b'}) 
\rho_b (\vec b - \vec {b'}).
\end{equation}
If we assume that the $\rho(\vec b)$ is given by Fourier Transform of
the form factor of the hadron, then $A_{ab} (b)$ is given by,
\begin{equation}
\label{aob}
A_{ab}(b)={{1}\over{(2\pi)^2}}\int d^2\vec{q}{\cal F}_a(q) {\cal F}_b(q) 
e^{i\vec{q}\cdot \vec{b}},
\end{equation}
where  ${\cal F}_{a,b}$  are the electromagnetic form factors 
of the colliding hadrons. For protons this is given by the dipole expression
\begin{equation}
\label{dipole}
{\cal F}_{prot}(q)=[{{\nu^2}\over{q^2+\nu^2}}]^2,
\end{equation}
with $\nu^2=0.71\ {\rm GeV}^2$. For photons a number of 
authors \cite{FLETCHER,SARC}, on the basis of Vector Meson Dominance, 
have assumed the same functional form as for pion, i.e. the pole expression 
\begin{equation}
\label{pole}
{\cal F}_{pion}(q)={{k_0^2}\over{q^2+k_0^2}},
\end{equation}
with $k_0 = 0.735$ GeV from the measured pion form factors,
 changing the value of the scale parameter $k_0$, if necessary 
in order to fit the data.

Yet another  philosophy would be to assume that  the b-space 
distribution of partons is the Fourier transform of the
transverse momentum distribution of the colliding system~\cite{BN}. 
To leading order, this transverse momentum
distribution can be  entirely due to an intrinsic
transverse momentum of partons in the parent hadron. While
the intrinsic transverse momentum ($k_T$) distribution of partons in a 
proton 
is normally taken to be  Gaussian,  a choice which can be justified in QCD
based models \cite{nak},  in the case 
of photon  the origin of all  partons can, in principle, be traced back to 
the hard vertex $\gamma \ q \bar q$.  Therefore, also in the intrinsic
transverse momentum philosophy,  one can expect the  $k_T$ distribution 
of photonic partons to be different from that of the partons in the proton. 
The expected functional dependence can be deduced  using the origin 
of photonic partons from the $\gamma \rightarrow q \bar q$ splitting. 
For the photon one can argue that the intrinsic 
transverse momentum ans\"atze would imply the use of a 
different value of the parameter $k_0$~\cite{rohinipl,tran}, which is 
extracted from data involving `resolved'~\cite{review} photon 
interactions~\cite{ZEUS}, in the pole expression for the form factor. 
By varying $k_0$ one can then explore  various possibilities, i.e. 
the ${\rm VMD}/\pi$ hypothesis if $k_{0}=0.735 $ GeV, or the intrinsic 
transverse distribution case for other values of $k_0$~\cite{ZEUS}.

The ansatz of eqs.~\ref{eik1}--~\ref{eik3} and \ref{numb}, requires
that the overlap function be normalised to unity, i.e.,
\be
\int d^2 \vec b A_{ab}(b) = 1.
\ee
Taking the form factor ansatz for the proton we then have
\begin{equation}
A_{pp}(b)={{\nu^2}\over{96 \pi}} (\nu b)^3{\cal K}_3(\nu b)\ \ \ \ 
\ \ \ \ A_{\gamma \gamma}={{k_0^3b}\over{4 \pi}}{\cal K}_1(k_0b)
\label{appgm}
\end{equation}
and 
\begin{equation}
A_{\gamma {\rm p}}(b)=
{{k_0^2\nu^4}\over{2\pi(\nu^2-k_0^2)^2}}
\left[{\cal K}_0(k_0b)-{\cal K}_0(\nu b)\right] 
+ {{k_0^2\nu^2}\over{4 \pi(k_0^2-\nu^2)}}\nu b {\cal K}_1(\nu b)
\label{agmp}
\end{equation}
where $\nu$ and $k_0$ are the scale factors mentioned earlier and
${\cal K}_{i}$ are the modified Bessel functions.

If we look at eqs.~\ref{siginel}-~\ref{numb} it is easy to see that $A_{ab}(b)$
and $P_{\rm ab}^{had}$ always appear in the combination 
$A_{ab}(b)/P_{ab}^{had}$~\cite{manuel,review}. Hence only one of 
them can be varied independently. Note also that $\sigma^{\rm soft}$
can always be renormalised since it is a function fitted to the low
energy data. 
By looking at eq.~\ref{numb} 
we can see that if the $s$-dependence of the $jet$ cross-sections
were similar for {\it all} the colliding particles, then the difference in the
$s$-dependece of the total/inelastic cross-sections can be estimated by 
looking at the behaviour of $A_{ab}$. It is also clear then that 
changing the scale parameter $k_0$  in $A_{ab}(b)$ is equivalent to 
changing $P_{ab}^{had}$. Note also that
\be
P_{\gamma {\rm p}}^{had} = P_\gamma ^{had};\;\;\; 
P_{\gamma \gamma}^{had} = (P_\gamma^{had})^2.
\ee 
Hence, in analysing the photon-induced reactions, i.e, the \gamp\
and \gamgam\ cross-sections, the only hadronisation probability 
that is an independent parameter is $\phad \equiv \pghad $.

Thus now we are ready to  list the total number of inputs on which the
EMM predictions depend:
\begin{itemize}
\item  The soft cross-sections $\sigma_{ab}^{\rm soft}$ ,
\item $p_{tmin}$ and the parton densities in the colliding hadrons,
\item \phad\ and the ansatz as well as the scale parameters for the
$A_{ab}(b)$,
\end{itemize}

Out of these the protonic and photonic parton densities are known from 
$eP$ and $e \gamma$ DIS. The nonperturbative part $\sigma_{\gamp}^{\rm soft},
\sigma_{\gamgam}^{\rm soft}$ has to be determined from some fits. We outline
the procedure used by us below. It is true that the jet cross-sections 
of eq.~\ref{miniold} depend very strongly on the value of $p_{tmin}$. Hence 
it would be useful to have an independent information of this parameter, 
which as 
said before separates the perturbative and nonperturbative contribution in an
ad hoc manner. Luckily, there is more direct evidence that the ansatz of
eq.~\ref{siginel} can describe some features of hadronic interactions.
Event generators which have built in multiple parton interactions
in a given $a b$ interaction, for the case of $a,b$ being $p, \bar p$,
were shown~\cite{torbzil} to explain many features of the hadronic interactions
such as multiplicity distributions with a $p_{tmin}$ around $1.5$ GeV. Recent
analyses of the \gamp\ interactions seem to show~\cite{seymor} again that 
a consistent description of many features such as energy flow, 
multiplicity distributions is possible with a $p_{tmin}$ value 
between $1.5$ -- $2.0$ GeV.

The energy dependence of $\sigma_{ab}^{\rm jet}$ as defined in 
eq.~\ref{miniold} will of course get reflected in the energy rise of 
the eikonalised total
or inelastic cross-section. It is therefore instructive to see how this depends
on the type of the colliding particles. We compare this for $p \bar p$,
\gamp\ and \gamgam\ case, where we have multiplied the \gamp\ and \gamgam\
jet cross-sections by factor of $\alpha$ and $\alpha^2$ respectively.
\begin{figure}
\begin{center}
\mbox{\epsfig{file=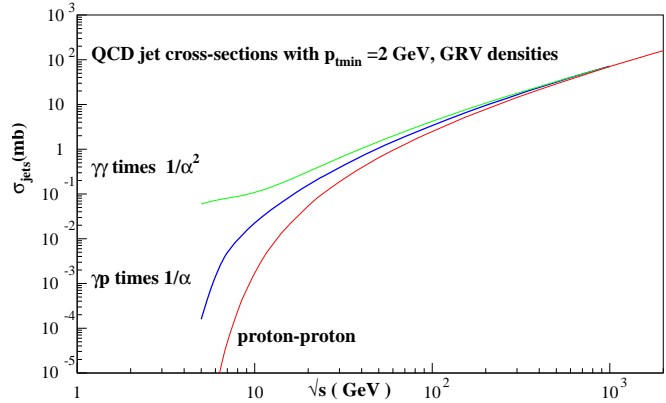,height=60mm}}
\end{center}
\caption{Energy dependence  of  $\sigma_{ab}^{\rm jet}$ for $p_{tmin} = 2 GeV$.
\label{figure2}}
\end{figure}
In the comparison of Fig.~\ref{figure2} we have used the GRV, LO 
parametrisations for both the proton~\cite{grvpr} and the photon~\cite{grvph}. 
We note here that at high energies, the rise with energy of the jet 
cross-sections is very similar in all the three cases, when difference
between a photon and a proton is accounted for.
A  study of  the $b$-dependence of the respective
$A_{ab}(b)$ given by eqs.~\ref{appgm},~\ref{agmp}, shows  that the photon
is much `smaller' as compared to the proton in the transverse space,
which is also understandable as the photon after all owes its 
`structure' to the hard $\gamma q \bar q$ vertex. Hence, we expect that
the damping of the cross-section rise due to multiple scattering 
for photons will be less than for a proton. This, coupled with the 
above observation  of the $jet$ cross-section, implies that in the EMM, 
total/inelastic cross-sections are expected to rise faster with energy
as we replace a proton by a photon.  I.e., an increase in the 
pomeron intercept as we go from $\bar p p$ to \gamp\ and \gamgam\, 
as indicated by the data, is expected in the EMM framework.

\section{Results}
Now in this section let us spell out our strategy of fixing the various inputs
to the EMM. We restrict our analysis only to the photon-induced
processes, i.e., \gamp\ and \gamgam\ cross-section. We follow the same 
procedure as we had adopted in~\cite{rohinipl}, i.e., we fix all the inputs to
the EMM  by a fit to the data on the available photoproduction data on
$\sigma_{\gamp}$. Here we do not include the data~\cite{Bernds,talk}, 
shown in Fig.~\ref{figure1}, which has been obtained by an extrapolation 
of the low $Q^2$ data to $0$. We determine $\sigma_{\gamp}^{\rm soft}$ by a 
fit to the photoproduction data using a form suggested in~\cite{SARC},
\begin{equation}
\label{soft}
\sigma^{\rm soft}_{\gamp} =\sigma^0_{\gamp} +
{{{\cal A}_{\gamp}}\over{\sqrt{s}}}+{{{\cal B}_{\gamp}}\over{s}}.
\end{equation}
We then determine ${\cal A}_{\gamp}, {\cal B}_{\gamp} $ and $\sigma^0_{\gamp}$
from the best fit to the low-energy photoproduction data, starting from the 
quark-model motivated ansatz $\sigma_{\gamp}^0 = 2/3 \sigma_{\bar p p}$.
In earlier work~\cite{rohinipl} we had used the inelastic formulation. Now we
have repeated the same exercise with the total cross-section formulation of 
the EMM,  which we believe is the more appropriate to use~\cite{block}. 
The results of our fit, using the total cross-section formulation of the EMM,
are shown 
\begin{figure}
\begin{center}
\mbox{\epsfig{file=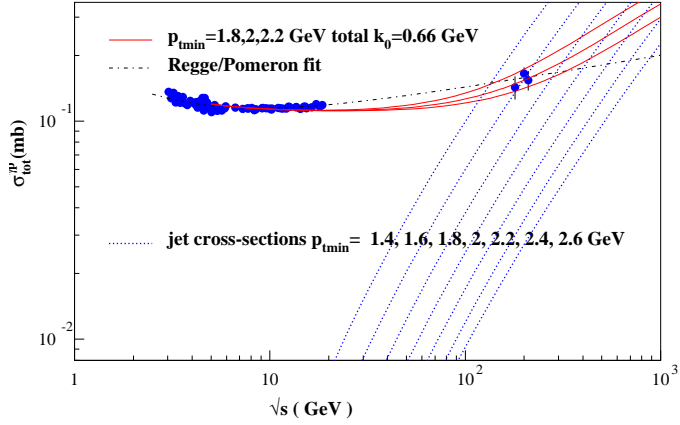,height=60mm}}
\end{center}
\caption{Comparison of the photoproduction data with the EMM fits with total
cross-section formulation, and jet cross-sections as a function of $p_{tmin}$.
\label{figure3}}
\end{figure}
in Fig.~\ref{figure3}. The fit values of the parameters are
\begin{equation}
\sigma^{0}_{\gamp} = 31.2\;\; {\rm mb}\;\;\;
;  {\cal A}_{\gamp} = 10\;\; {\rm  mb} \; {\rm GeV}; \;\;\;  
{\cal B}_{\gamp}  = 37.9 \;\; {\rm mb}\;\; {\rm GeV}^2 . 
\end{equation}
As compared with the similar exercise done in~\cite{rohinipl}, we find that
the rise of the  eikonalised cross-sections with \rts\ is faster in this
case than in the inelastic formulation. However, $p_{tmin} = 2$ GeV is still 
the best value to use, as seen from Fig.~\ref{figure3}. We use here the form 
factor ansatz for the proton and   the intrinsic $k_T$ ansatz for the
photon with a value of parameter $k_0 = 0.66 $ GeV,  which
corresponds to the central value from  the measurement~\cite{ZEUS} of the
intrinsic $k_T$ distribution. We have  used GRV distributions for both the
proton and photon and $\phad = 1/240$. We also find, similar
to the analysis in the inelastic formulation by us~\cite{rohinipl} and 
others~\cite{SARC,forshaw}, that the description of the photoproduction data 
in terms of a single eikonal leaves leeway for improvement. We restrict
ourselves to the use of a single eikonal, so as to minimize our parameters 
but note that  this can perhaps be cured by using an energy dependent \phad\
or alternatively an energy dependent $k_0$.

Now, having fixed all the inputs for the \gamp\ case,
we determine the corresponding parameters for the \gamgam\ case again
by appealing to the Quark Model considerations and we use,
\be
\sigma^{soft}_{\gamgam} = {2\over 3} \sigma^{soft}_{\gamp}.
\ee
All the other inputs are exactly the same as in the \gamp\ case.
In this manner, we have really {\it extrapolated} our results from
\gamp\ case to the \gamgam\ case. The results of our extrapolation 
are shown
\begin{figure}
\begin{center}
\mbox{\epsfig{file=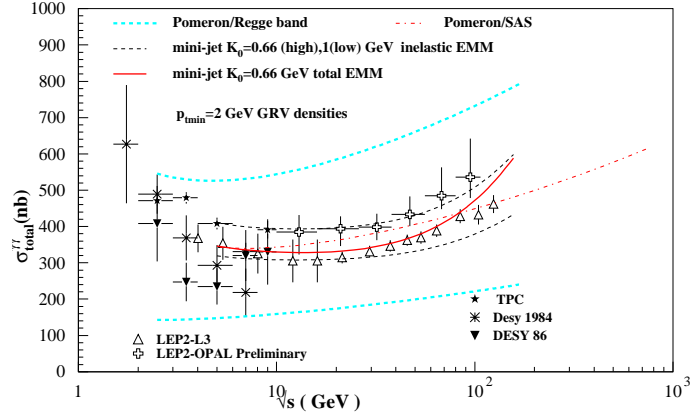,height=60mm}}
\end{center}
\caption{EMM  predictions for \gamgam\ cross-sections for total
and inelastic cross-section formulations along with data and
Regge -Pomeron prediction.
\label{figure4}}
\end{figure}
in Fig.~\ref{figure4}.

Notice that the overall normalization of the photonic cross-section depends
upon $P^{had}$. When extrapolating from photoproduction
to photon-photon using the inelastic assumption, we had used
$P^{had}=1/200$, which can be thought of as corresponding
 to a 20\% non-VMD component.
Using the total cross-section formulation, the low energy production data 
suggest rather to use $P^{had}=1/240$, a value which implies that the 
photon is
practically purely a vector meson. Then, Fig.\ref{figure4} shows that
in the total cross-section formulation,
the extrapolation from \gamp\ to $\gamma \gamma$ leads to 
 cross-section which lies lower at low energies, but rises faster,
than in the inelastic fits,  for the same values of the parameters. 

Both the inelastic and the total cross-section are seen to rise faster than
 is
expected in an universal pomeron picture. This feature is same both for
\gamp\ and \gamgam\ cases. We show the dependence of our results on the
scale parameter $k_0$.
 The band in the figure corresponds to using the Regge-Pomeron 
hypothesis of eq.~\ref{regpom}, measured values of $X_{ab},Y_{ab}$ 
for $\bar p p/pp$, \gamp\ case and the  factorisation  idea~\cite{terezawa}
$$
X_{\gamgam} = X^2_{\gamp}/X_{p p};\;\;
Y_{\gamgam} = Y^2_{\gamp}/Y_{ p p}.
$$
Here $X(Y)_{pp}$ stand for an average for $pp$ and $\bar p p$ case.

We see that while our analysis using inelastic formulation and the default 
value of $k_0 = 0.66$ GeV~\cite{ZEUS} gave predictions
closer to the OPAL data  the total cross-section formulation,\ {\it for the
same value of $k_0$}, gives results closer to the L3 data, as already 
pointed out in \cite{block}. The 
sensitivity of the predictions to the difference between different 
parametrisations for the photonic partons increases with energy. At higher
energies one is sensitive to the low-$x$ region about which not much is known.
Our earlier analysis~\cite{rohinipl} in the inelastic formulation had shown 
that the \gamgam\ cross-sections rise more slowly for the SAS~\cite{SAS}
parametrisation of the photonic parton densities. The dependence of
our results in this analysis on the parton densities in the photon will
be presented elsewhere~\cite{twous}.

\begin{figure}
\begin{center}
\mbox{\epsfig{file=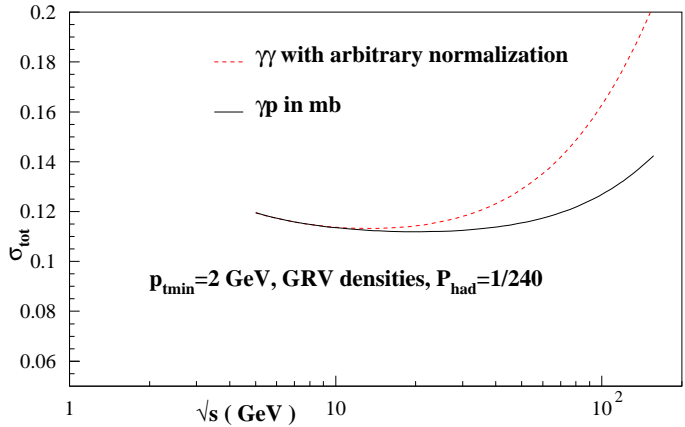,height=60mm}}
\end{center}
\caption{Comparison of the energy dependence of the EMM  predictions 
for the total \gamgam\ and \gamp\ cross-sections 
with  $p_{tmin} = 2$ GeV, \phad\ = 1/240.
\label{figure5}}
\end{figure}
Fig.~\ref{figure5} shows a comparison of the eikonalised \gamp\ and \gamgam\
cross-sections, in the total formulation, with the different parameters for
\gamp\ and \gamgam\ case  related as described before. We see indeed that in the EMM the \gamgam\ cross-sections rise faster with \rts\ than \gamp\ case, as 
was expected from the results shown in Fig.~\ref{figure3} and arguments following that. However, the dependence of this observation on \phad\ and/or the scale
parameter needs to be still explored.

\section{Conclusions}

In conclusion  we discuss the results of an analysis of total and 
inelastic hadronic cross-sections for photon induced processes
in the framework  of an eikonalised minijet model (EMM).  We have fixed
various input parameters to the EMM calculations  by using the data
on photoproduction cross-sections and then made predictions for
\siggmhad\ using {\it same} values of the  parameters. We then compare
our predictions with the recent measurements of \siggmhad\ from LEP.
We find that the total cross-section formulation of the EMM predicts
faster rise with \rts\ as compared to the inelastic one, for the {\it
same} value of $p_{tmin}$ and scale parameter $k_0$. In the former case 
our extrapolations yield results closer to the L3 data whereas in the
latter case they are closer to the OPAL results. We also find that
in the framework of EMM it is natural to expect a faster rise with
\rts\ for the \gamgam\ case as compared to the \gamp\ case.
\section{Acknowledgement}
It is a pleasure to thank Goran Jarlskog and Torbjorn Sj\"ostrand for 
organising this workshop which provided such a pleasant atmosphere
for very useful discussions. G.P.  is grateful to Martin Block for
clarifying  discussions on the total cross-section formulation.  
R.M.G. wishes to acknowledge support from  Department of Science and Technology
(India) and National Science Foundation,under NSF-grant-Int-9602567
and G.P from the EEC-TMR-00169.

\end{document}